\documentstyle[aps,twocolumn]{revtex}
\begin{document}
\title{
Quantum Cloning of Mixed States in Symmetric Subspace
}

\author{Heng Fan}
\address{
Quantum computation and information project,
ERATO, Japan Science and Technology Corporation,\\
Daini Hongo White Building 201, Hongo 5-28-3, Bunkyo-ku, 
Tokyo 113-0033, Japan.}
\maketitle
                                                        
\begin{abstract}
Quantum cloning machine for
arbitrary mixed states in symmetric subspace
is proposed. 
This quantum cloning machine
can be used to copy part of the output state
of another quantum cloning machine and is useful
in quantum computation and quantum information.  
The shrinking factor of this quantum cloning
achieves the well-known upper bound.
When the input is identical pure states,
two different fidelities of this cloning machine 
are optimal.
\end{abstract}
       

A quantum state cannot be cloned exactly because of the
no-cloning theorem\cite{WZ}. However, quantum cloning
approximately (or probablisticaly)
is necessary in quantum computation and quantum information\cite{NC}.
Suppose we have the following task: we have a
pure unknown quantum state
in 2-level system (qubit) $|\Psi \rangle  $. 
We need one copy of this quantum state
to perform one quantum computation.
But we do not need it to be exactly the original one.
A copy of $|\Psi \rangle $ with fidelity of at least $7/9$
can give a reliable result. 
And also we need another 3 identical quantum states each with the
fidelity of at least $79/108$ to perform another reliable quantum computation.
If the fidelities of the quantum states are less than the 
demanding fidelities, the quantum computation will not
be reliable any more. This is certainly a simple and rather practical
quantum cloning task. However, we still cannot reach this simple
goal by the present available optimal quantum cloning machines. 

Let's next analyze why the
present available quantum cloning machines fail to do this work. 
1, First we try to use the 1 to 4 optimal quantum cloning machine
proposed by Gisin and Massar\cite{GM} to do this work.
By this cloning machine, we can copy $|\Psi \rangle $
to 4 identical quantum states
each with fidelity $3/4$ which is larger than $79/108$ but
less than $7/9$. That means we can obtain a reliable result
in the second quantum computation but
we cannot have a reliable result in the first quantum computation. 
2, We may use first the 1 to 2 cloning machine which is proposed
by Buzek and Hillery\cite{BH,BH1}. With one output state doing
the first quantum computation, then we use 
another quantum state as input and use the
1 to 3 Gisin-Massar
cloning machine to create another 3 identical quantum states.
One can find the quantum state of the first cloning machine
can achieve the fidelity $5/6$ which is better than 
the demanding fidelity $7/9$. However, the 3 identical quantum
states can only achieve the fidelity $37/54$ which is
lower than the demanding fidelity. So, we cannot finish
our task by using this method.   
3, One is perhaps tempted to use Cerf's asymmetric quantum cloning
machine\cite{Cerf} to do this work. 
The advantage of using Cerf's asymmetric cloning machine is 
that we can let one quantum state achieve the fidelity
$7/9$ while another one still has the optimal fidelty
since this cloning machine achieve the bound of the
no-cloning theorem proposed by Cerf. Then we use the
Gisin-Massar 1 to 3 cloning machine to create another
3 identical quantum states. By calculation we can 
find if one quantum state has fidelity $7/9$, 
the optimal fidelity achieved by another quantum state
is $11+2\sqrt{6}/18\approx 0.88$. However, we need it
at least $11/12\approx 0.92$ to create reliable 3 identical quantum states
by using Gisin-Massar cloning machine. So, we still cannot achieve
our aim.

Is this task in principal cannot be accomplished since no-cloning
theorem?
By simple calculation we can show that this goal in principal
can be achieved. The following is one method.
We can first use the Gisin-Massar
1 to 3 cloning machine. And after this quantum cloning,
we can use one quantum state which
has fidelity $7/9$ to perform the first quantum computation
which will give the reliable result.
The remaining quantum state is a two qubits mixed state
in symmetric subspace. The theory of 
Bruss, Ekert and  Macchiavello (BEM) \cite{BEM} shows that the optimal
shrinking factor, which has a simple relation with fidelity
for pure state, of 2 to 3 cloning machine can achieve $5/6$.
So, we can obtain 3 quantum states each with fidelity     
$79/108$ which will also give a reliable result
in the second quantum computation.
Thus both two quantum computations will give 
reliable results. 
The Gisin-Massar cloning machine can only copy
2 {\it identical ~pure} states to 3 copies. 
The problem is that here the input
state of 2 qubits is a mixed states in symmetric space which
cannot be copied by the available cloning machines.
So we must
construct a 2 to 3 optimal cloning machine
which can use a mixed state in symmetric subspace as input.

One can imagine that a lot of other similar problems
exist in the quantum computation.
In this paper we will first present explicitly
the quantum cloning machine
which can accomplish the above mentioned task.
And further some more complicated tasks similar to the
above one can also be accomplished.
Our result is actually rather general.
We will present the optimal quantum cloning
machine which can use arbitrary d-level mixed states in symmetric
subspace as input. And the cloning machine is optimal since
it achieves the upper bound of the shrinking factor \cite{BEM,W,KW}.     
The cloning machine presented in this paper can also be 
used directly to analyze the security of quantum key distribution
when all $d+1$ mutually unbiased states are used which
provides the most secure protocol in d-level quantum system, here
$d$ is assumed to be prime number in this quantum key distribution
protocol. 
 
Let's study the quantum cloning task presented above.
We assume the available unknow quantum state is expressed
as $|\Psi \rangle =\alpha |\uparrow \rangle +\beta |\downarrow
\rangle $
with $|\alpha |^2+|\beta |^2=1$. First we use Gisin-Massar
1 to 3 cloning machine\cite{GM} which will create 3 copies
of $|\Psi \rangle $ approximately.
One output state is $\rho _{red.}=5/9|\Psi \rangle
\langle \Psi |+2/9\cdot I$, where $I$ is the identity.
The fidelity of $\rho _{red.}$ with the original
quantum state $|\Psi \rangle $ is $7/9$ which achieve
the threshold to give a reliable result in the
first quantum computation.
The remaining 2 qubits state which is obtained
after tracing out 1 qubit used for the first
quantum computation from the output state takes the form
\begin{eqnarray}
\rho _{red.}^{(2)}&=&x_0|2\uparrow \rangle
\langle 2\uparrow |
+x_1|2\uparrow \rangle \langle
\uparrow ,\downarrow |
\nonumber \\
&+&x_1^*|\uparrow ,\downarrow \rangle
\langle 2\uparrow |+
\frac {1}{3}|\uparrow ,\downarrow \rangle \langle 
\uparrow ,\downarrow |
+x_1|\uparrow ,\downarrow \rangle \langle
2\downarrow |\nonumber \\
&+&x_1^*|2\downarrow \rangle \langle
\uparrow ,\downarrow |
+x_2
|2\downarrow \rangle \langle 2\downarrow|,   
\label{2input}
\end{eqnarray}
where $x_0=(\frac {1}{18}+\frac {5}{9}|\alpha |^2)$,
$x_1=\frac {5\sqrt{2}}{18}\alpha \beta ^*$,
$x_2=\frac {1}{18}+\frac {5}{9}|\beta |^2$, and $|2\uparrow \equiv 
|\uparrow \uparrow \rangle $, similar for 
$|2\downarrow \rangle $,
$|\uparrow ,\downarrow \rangle =(|\uparrow \downarrow \rangle
+|\downarrow \uparrow \rangle )/\sqrt{2}$ is the normalized
symmetric state. This state is a mixed state so we cannot
use available quantum cloning machines to obtain
3 copies. We propose the following 2 to 3 cloning machine
to accomplish this task
\begin{eqnarray}
U|2\uparrow \rangle \otimes R
&=&\frac {\sqrt{3}}{2}|3\uparrow \rangle \otimes R_{\uparrow }
+\frac {1}{2}|2\uparrow ,\downarrow \rangle \otimes R_{\downarrow },
\label{GM1}\\
U|2\downarrow \rangle \otimes R
&=&\frac {1}{2}|\uparrow ,2\downarrow \rangle \otimes R_{\uparrow }
+\frac {\sqrt{3}}{2}|\downarrow \rangle \otimes R_{\downarrow },
\label{GM2}\\
U|\uparrow ,\downarrow \rangle \otimes R
&=&\frac {1}{\sqrt{2}}|2\uparrow ,\downarrow \rangle \otimes R_{\uparrow }
+\frac {1}{\sqrt{2}}|\uparrow ,2\downarrow\rangle \otimes R_{\downarrow },
\label{GM}
\end{eqnarray}
where state $|2\uparrow ,\downarrow \rangle $ is a normalized 
symmetric state with 2 spin up and 1 spin down.
We remark that the quantum cloning transformations (\ref{GM1},\ref{GM2})
are the same as Gisin-Massar's original cloning machine where the input
are identical pure states. The cloning transformation 
(\ref{GM}) is a new relation. 
In the quantum cloning processing with input $\rho _{red.}^{(2)}$ in 
(\ref{2input}), we first add blank state and
the ancilla state, perform the quantum cloning transformations
listed as (\ref{GM1},\ref{GM2},\ref{GM}), trace out the
ancilla states and finally obtain 3 identical copies.
The final reduced density operator of one single 
qubit has fidelity $79/108$ compared with the initial available qubit 
$|\Psi \rangle $. Thus we show explicitly how to
accomplish the simple but rather practical cloning task in
quantum computation.
This is a simple example to show that the cloning machine
which can use any mixed states in symmetric space is
very important in quantum computation. 

A much general problem is that part of the output
state from one cloning machine is used for one quantum computation.
The remaining quantum state need to be further cloned to
create more copies but with a lower fidelity
for another quantum computation.
One can easily imagine more complicated
problems where more than 2 cloning machines are needed. 
All of these problems can
be solved if we can construct the cloning machines which
can copy any mixed states in symmetric space.  
 
With one simple but non-trivial example solved, we next
present our general result.   
We assume $\{ |i\rangle ,i=1,\cdots,d \}$ as orthonormal basis of d-level
quantum system. We define
$\vec{m}=\{ m_1,\cdots ,m_d\}$, 
and denote $|\vec{m}\rangle $
as completely symmetrical states with $m_i$ states in
$|i\rangle $, where $\sum _{i=1}^dm_i=M$, these are orthonormal
basis of symmetric subspace of $M$ d-level quantum system.  
Any quantum states in symmetric subspace is invariant under
arbitrary permutations.
An arbitrary quantum state, mixed and/or entangled, in symmetric
subspace can be written as
\begin{eqnarray}
\rho =\sum _{\vec{m},\vec{m}'}\lambda _{\vec{m},\vec{m}'}
|\vec{m}\rangle \langle \vec{m}'| .
\label{input}
\end{eqnarray} 
The dimension of this matrix is $(M+d-1)!/M!(d-1)!$, and
we have trace normalization $\sum _{\vec{m}}
\lambda _{\vec{m},\vec{m}}=1$.
One example of this matrix for 2-level system is
$\rho _{red.}^{(2)}$ presented in (\ref{2input}).
Since the output state of most of the present quantum cloning
machines is symmetric state. Part output state is still
in symmetric subspace, so can be represented as the form (\ref{input}). 
For example, the Gisin-Massar 1 to 3 cloning machine creates
3 copies. And 2 copies in the output take
the form $\rho _{red.}^{(2)}$ in (\ref{2input}).
By tracing out $M-1$ d-level quantum system, we can obtain
the reduced density operator of a single d-level quantum state.
After some calculations, we find the reduecd density
operator takes the form
\begin{eqnarray}
\rho _{red.}&=&\sum _i|i\rangle \langle i|
\sum _{\vec{m},\vec{m}'}\lambda _{\vec{m},\vec{m}'}\frac {m_i}{M}
\delta _{\vec{m},\vec{m}'}
\nonumber \\
&&+\sum _{i\not= j}|i\rangle \langle j|
\sum _{\vec{m},\vec{m}'}\lambda _{\vec{m},\vec{m}'}
\frac {\sqrt{m_i(m_j+1)}}{M}
\nonumber \\
&&\delta _{m_i,m_i'+1}\delta _{m_j+1,m_j'}
\delta '_{\vec{m},\vec{m}'},
\label{redinput}
\end{eqnarray}
where $\delta _{\vec{m},\vec{m}'}\equiv \delta _{m_1,m_1'}
\cdots \delta _{m_d,m_d'}$ is a series of Kronecker delta functions,
in the second term of r.h.s., $\delta '_{\vec{m},\vec{m}'}$ is similar
to $\delta _{\vec{m},\vec{m}'}$ but without the terms $i$ and $j$.
Obviously, we have $M$ equivalent reduced density operators 
as form (\ref{redinput}).
As an example, there are 2 identical reduced density operators 
from $\rho _{red.}^{(2)}$ in (\ref{2input}) with the form 
$\rho _{red.}=5/9|\Psi \rangle
\langle \Psi |+2/9\cdot I$ as already presented.
Suppose we want to copy the quantum state $\rho $ so that
we can obtain $N$ $(N\ge M)$ equivalent reduced density operators
$\rho ^{(out)}_{red.}$. 
Our aim is to keep $\rho _{red.}$ as invariance as
possible. The result in Ref.\cite{BEM} shows, if we want the
quality of the copies is independent of the input, we should have
the following relation,
\begin{eqnarray}
\rho _{red.}^{(out)}=\frac {1-f}{d}I
+f\rho _{red.},
\label{scalar}
\end{eqnarray}
where $I$ is identity, i.e., completely mixed state in d-level quantum
system acting as the noise, $f$ is the shrinking 
factor ranging from 0 to 1. 
Apparently, the larger $f$ is, the better quality the copies have.    
We remark that the shrinking factor and the fidelity have a simple linear
relation if $\rho _{red.}$ is a pure state.
BEM bound shows that the optimal shrinking factor takes the value
$f=M(N+d)/N(M+d)$ which is independent of the input state. 
The purpose of this paper is to find
the cloning transformation achieving this bound.

We propose the following universal quantum cloning machine,
\begin{eqnarray}
U|\vec{m}\rangle \otimes R
=\sum _{\vec{k}}\alpha _{\vec{m}\vec{k}}
|\vec{m}+\vec{k}\rangle \otimes R_{\vec{k}},
\label{clone}
\end{eqnarray}
where $\sum _{\vec{k}}$ means summation over all possible parameters
under the restriction $\sum _ik_i=N-M$, $R$ represents blank states
and ancilla states of the cloning machine before the cloning
processing. $R_{\vec{k}}$ are the ancilla states of the output,
we can simplely realize them by symmetric quantum state $|\vec{k}\rangle $.
A simple example of this quantum cloning machine has already been
presented in (\ref{GM1},\ref{GM2},\ref{GM}).
Performing unitary transformation $U$,
tracing out the ancilla states, we can obtain the output density
operator 
$\rho ^{(out)}=Tr_aU(\rho \otimes R)U^{\dagger }$.
The amplitude of the output states take the form
\begin{eqnarray}
\alpha _{\vec{m}\vec{k}}=\eta 
\sqrt{\prod _i\frac {(m_i+k_i)!}{m_i!k_i!}},
\label{parameter}
\end{eqnarray}
where $\eta =\sqrt{(N-M)!(M+d-1)!/(N+d-1)!}$ is the normalization factor.
Our aim is to copy the input state $\rho $ in (\ref{input}).
By using the cloning machine (\ref{clone}), we find the output
density operator is the following,
\begin{eqnarray}
\rho ^{(out)}=
\sum _{\vec{m},\vec{m}'}\sum _{\vec {k}}
\lambda _{\vec{m},\vec{m}'}\alpha _{\vec{m}\vec{k}}
\alpha _{\vec{m}'\vec{k}}|\vec{m}+\vec{k}\rangle
\langle \vec{m}'+\vec{k}|.
\label{output}
\end{eqnarray}  
We remark that the quality of the copies is quantified by
the shrinking factor between input and output
of a single d-level reduced density operators.
With the help of the result in (\ref{redinput}),
the reduced density operator of the output (\ref{output}) can be written
as
\begin{eqnarray}
&&\rho ^{(out)}_{red.}
=\sum _i|i\rangle \langle i|\sum _{\vec{m},\vec{k}}
\lambda _{\vec{m},\vec{m}}\alpha ^2_{\vec{m}\vec{k}}
\frac {m_i+k_i}{N}
\nonumber \\
&&+\sum _{i\not =j}|i\rangle \langle j|
\sum _{\vec{m},\vec{m}',\vec{k}}\lambda _{\vec{m},\vec{m}'}
\alpha _{\vec{m}\vec{k}}\alpha _{\vec{m}'\vec{k}}
\nonumber \\
&&\frac {\sqrt{(m_i+k_i)(m_j+k_j+1)}}{N}
\delta _{m_i,m_i'+1}\delta _{m_j+1,m_j'}\delta '_{\vec{m},\vec{m}'}.
\label{redoutput}
\end{eqnarray}
Initially, we have $M$ identical reduced density operators (\ref{redinput}),
by using the quantum cloning machine (\ref{clone}), we now obtain
$N$ equivalent reduced density opertors (\ref{redoutput}).
Let's go back to the example again to see this quantum
cloning processing. Initially we have 2 identical quantum
states $\rho _{red.}=5/9|\Psi \rangle \langle \Psi |
+2/9\cdot I$ in (\ref{2input}). After the cloning
processing, we obtain 3 identical quantum states
each takes the form
$\rho _{red.}^{(out)}=25/54|\Psi \rangle \langle \Psi |
+29/108\cdot I$. The fidelity of this state with
$|\Psi \rangle $ is $79/108$ as already presented.

To show the cloning machine (\ref{clone}) 
is universal and optimal, we need to show
the input and out reduced density opertors (\ref{redinput}) and 
(\ref{redoutput})
satisfy the relation (\ref{scalar}) and the shrinking factor achieves
the BEM bound. 
Let's first check the off-diagonal elements of (\ref{redoutput}).
Substitute the results 
presented in (\ref{parameter}) into (\ref{redoutput}), and with
the help of the relation 
\begin{eqnarray}
&&\sum _{\vec{k}}^{N-M}\prod _{l=1}^d
\frac {(m_l+k_l)!}{k_l!m_l!}(m_j+k_j+1)
\nonumber \\
&=&(m_j+1)\frac {(N+d)!}{(N-M)!(M+d)!},
\end{eqnarray}
we find relation (\ref{scalar}) is satisfied and the optimal shrinking factor
is achieved. The diagonal elements can be similarly checked with
the help of the trace normalization relation.
Thus we show the cloning transformation (\ref{clone}) with 
amplitude parameters (\ref{parameter}) is the
universal quantum cloning machine and arbitrary symmetric
input states can be cloned optimally in the sense BEM bound
is achieved. 
Besides the usefulness of this cloning machine
for the quantum cloning task,
the result presented here is also interesting.
We know
the BEM bound can be achieved for
identical pure input states by 
Werner\cite{W} and Keyl and Werner\cite{KW} 
(WKW) cloning machine, see also \cite{Z}.
For arbitrary states in symmetric subspace which include
the identical pure states as a special case, intuitively,
this bound cannot be achieved since compared with
pure states case, we know less about the input.
But the result in this paper shows that 
we can copy arbitrary symmetric states 
as good as we copy identical pure states.
And the cloning
transformation achieving the BEM bound is explicitly given.  
Thus BEM bound is the tight bound not only for
pure states but also for arbitrary states in symmetric
subspace.          

Next, we shall show the cloning transformation (\ref{clone},\ref{parameter})
realizes the WKW \cite{W,KW} cloning machine.
WKW cloning machine can copy optimally $M$ identical pure states
to $N$ copies. The quality of the copies is quantified by 
both fidelity of redueced density operators of the single quantum
state between input and output and
the fidelity between the whole output and the perfect cloning output.  
We will use the cloning transformation presented in this paper
to copy $M$ identical pure states to $N$ copies and show 
both optimal fidelities
are achieved.
An arbitrary pure state in d-level system can be written as
$|\Psi \rangle =\sum _ix_i|i\rangle $, where $\sum _i|x_i|^2=1$.
$M$ identical pure input states lie in 
symmetric subspace of form (\ref{input}).
Explicitly it can be written as
\begin{eqnarray}
|\Psi \rangle ^{\otimes M}\equiv |M\Psi \rangle 
=\sqrt{M!}\sum _{\vec {m}}
\prod _{i}\frac {x_i^{m_i}}{\sqrt{m_i!}}|\vec{m} \rangle .
\end{eqnarray}
Substitute this input into the cloning transformation (\ref{clone}),
we can find the output state density operator after tracing out
the ancilla states as
\begin{eqnarray}
\rho ^{(out)}&=&
M!\sum _{\vec{m},\vec{m}',\vec{k}}
\prod _i\frac {x_i^{m_i}(x_i^*)^{m_i'}}{\sqrt{m_i!m_i'!}}
\alpha _{\vec{m}\vec{k}}\alpha _{\vec{m}'\vec{k}}
\nonumber \\
&&|\vec{m}+\vec{k}\rangle \langle \vec{m}'+\vec{k}|.
\label{pureout}
\end{eqnarray}
This is a special form of output (\ref{output}).
The reduced density operators satisfy (\ref{scalar}) with 
input reduced density operator being simply 
pure state $|\Psi \rangle \langle \Psi |$.
The fidelity for single quantum state between
input and output thus can be obtained from the
optimal shrinking factor
\begin{eqnarray}
F&=&\langle \Psi |
\rho _{red.}^{(out)}|\Psi \rangle
=\frac {(d-1)f+1}{d}
\nonumber \\
&=&\frac {M(N+d)+N-M}{N(M+d)}.
\label{fidelity1}
\end{eqnarray}
By direct calculation,
we can also obtain the fidelity between the whole output (\ref{pureout}) and
the ideal copies
\begin{eqnarray}
\tilde {F}
=^{\otimes N}\langle \Psi |\rho ^{(out)}|\Psi \rangle ^{\otimes N}
=\frac {N!(M+d-1)!}{M!(N+d-1)!}.
\label{fidelity2}
\end{eqnarray}
Thus we show the cloning machine (\ref{clone},\ref{parameter})
can copy identical pure states optimally in the sense both
optimal fidelities (\ref{fidelity1},\ref{fidelity2}) 
can be achieved\cite{W,KW}.
So we give a realization of WKW cloning machine.  

Let $|\Psi \rangle $ be an arbitrary pure state 
in d-level quantum system, let $|\Psi _i^{\perp }\rangle ,i=1,\cdots d-1$ 
be orthonormal states which are also orthogonal to $|\Psi \rangle $.
The quantum cloning machine (\ref{clone},\ref{parameter})
can be reformulated as the following form
\begin{eqnarray}
&&U|m_1\Psi ,m_2\Psi _1^{\perp },\cdots, m_d\Psi _{d-1}^{\perp }\rangle 
\otimes R
\nonumber \\
&=&\sum _{\vec {k}}\alpha _{\vec {m}\vec {k}}
|(m_1+k_1)\Psi ,\cdots, (m_d+k_d)\Psi _{d-1}^{\perp }\rangle
\nonumber \\
&&~~~\otimes R_{\vec {k}}(\Psi ),  
\label{general}
\end{eqnarray} 
where ancilla states can also be realized in
symmetric subspace $R_{\vec {k}}(\Psi )=|k_1\Psi ,\cdots, k_d\Psi _{d-1}^{\perp }
\rangle $. 
If the input is $M$ identical pure states $|\Psi \rangle $, 
where $\vec {m}=\{ M,0,\cdots 0\} $,
this formula
gives a much simplified result since we do not need to expand the input
in terms of symmetric basis. And the two fidelities can be shown
to be optimal and thus provide a realization of WKW cloning machine.
For 2-level quantum system,
we recover the well-known optimal Gisin-Massar \cite{GM} universal
cloning machine.

A direct application of the cloning machine is to analyze the
security of the quantum key distribution. It is known when
$d$ is a prime number, there are $d+1$ mutually unbiased bases.
Besides the standard basis $\{ |i\rangle , i=1,\cdots d\} $,
other $d$ mutually unbiased bases can be written as
\begin{eqnarray}
|\psi _t^k\rangle 
=\frac {1}{\sqrt{d}}
\sum _{j=1}^d(\omega ^t)^{(d-j)}(\omega ^{-k})^{s_j}|j\rangle ,
t=1, \cdots, d, 
\end{eqnarray}
where $s_j=j+\cdots +(d-1),\omega =e ^{2\pi i/d}$. In quantum key distribution,
since all mutually unbiased bases are used, what eavesdropper
can do is to use optimal universal quantum cloning machine
to clone the sended quantum state to two copies.
Thus the clone machine (\ref{general}) can be used directly.
Explicitly, we have $|i\rangle \rightarrow
1/\sqrt{2(d+1)}\sum _{j=1}^d(|ij\rangle +|ji\rangle )
\otimes R_j$ and
\begin{eqnarray}
U|\psi _t^k\rangle \otimes R
&=&\sqrt{\frac{2}{d+1}}|2\psi _t^k\rangle \otimes R(\psi _t^k)   
\nonumber \\
&&+\frac {1}{\sqrt {d+1}}
\sum _{t'\not =t}|\psi _t^k,\psi _{t'}^k\rangle
\otimes R(\psi _{t'}^k).
\label{result}
\end{eqnarray}
This relation shows copied result explicitly for all input states.
The fidelity of the eavesdropper and the receiver's
quantum state is $F=(d+3)/2(d+1)$. So for individual
attack, the receiver can at most tolerate error rate
$1-F=(d-1)/2(d+1)$. 

Finally, we remark that the symmetric basis can be mapped to 
Fock states
represented by canonical boson operators up to a whole normalization 
factor as,
$|\vec{m}\rangle 
\rightarrow \prod _{i=1}^d
\frac {(a_i^{\dagger })^{m_i}}{\sqrt{m_i}}|vac\rangle $,
where $|vac\rangle $ is the vacuum state. The
Hamiltonian which can realize the optimal cloning
machine presented in (\ref{clone},\ref{parameter}),(\ref{general})
takes the form
$H=\sum _i(a_i^{\dagger }b_i^{\dagger }+a_ib_i)$.
The detailed results can be found in Ref.\cite{FWMI}. 

In conclusion, we give explicitly the quantum cloning transformation
which can copy arbitrary states in symmetric subspace.
This cloning machine can accomplish various quantum cloning
tasks in quantum information processing.

{\it Acknowlegements}:
The author would like to thank J.Kempe, H. Kobayashi, 
K.Matsumoto, B.S.Shi, 
X.B.Wang, G.Weihs, R.Werner, P.Zanardi and members of ERATO project 
for useful discussions.

\end{document}